\documentclass[structabstract]{aa}  
\usepackage{graphicx}
\usepackage{txfonts}
\usepackage{longtable,lscape,rotating}
\usepackage{multirow}
\usepackage{array}
\usepackage{color}
\usepackage{natbib}
\bibpunct{(}{)}{;}{a}{}{,} 
\begin{document}
   \title{Solar wind charge exchange X-ray emission from Mars}

   \subtitle{Model and data comparison}

   \author{D. Koutroumpa\inst{1}
          \and
          R. Modolo\inst{1}
          \and
          G. Chanteur\inst{2}
          \and
          J.-Y., Chaufray\inst{1}
          \and
          V. Kharchenko\inst{3}
          \and
          R. Lallement\inst{4}
                   }

   \institute{Universit\'e Versailles St-Quentin; CNRS/INSU, LATMOS-IPSL , 11 Bd d'Alembert, 78280, Guyancourt, France\\
              \email{dimitra.koutroumpa@latmos.ipsl.fr}
         \and
             Laboratoire de Physique des Plasmas, Ecole Polytechnique, UPMC, CNRS, Palaiseau, France
          \and
             Physics Department, University of Connecticut, Storrs, Connecticut, USA
          \and
             GEPI, Observatoire de Paris, CNRS, 92195, Meudon, France\\
              }

   \date{Received ; accepted }

 
  \abstract
   {}
   {We study the soft X-ray emission induced by charge exchange (CX) collisions between solar-wind, highly charged ions and neutral atoms of the Martian exosphere.}
   {A 3D multi species hybrid simulation model with improved spatial resolution (130~km) is used to describe the interaction between the solar wind and the Martian neutrals. We calculated velocity and density distributions of the solar wind plasma in the Martian environment with realistic planetary ions description, using spherically symmetric exospheric H and O profiles. Following that, a 3D test-particle model was developed to compute the X-ray emission produced by CX collisions between neutrals and solar wind minor ions. The model results are compared to XMM-Newton observations of Mars.}
   {We calculate projected X-ray emission maps for the XMM-Newton observing conditions and demonstrate how the X-ray emission reflects the Martian electromagnetic structure in accordance with the observed X-ray images. Our maps confirm that X-ray images are a powerful tool for the study of solar wind - planetary interfaces. However, the simulation results reveal several quantitative discrepancies compared to the observations. Typical solar wind and neutral coronae conditions corresponding to the 2003 observation period of Mars cannot reproduce the high luminosity or the corresponding very extended halo observed with XMM-Newton. Potential explanations of these discrepancies are discussed.}
   {}

   \keywords{atomic processes --
                solar wind --
                Planets and satellites: atmospheres --
                Planets and satellites: individual: Mars --
                X-rays: individuals: Mars
               }

   \maketitle
%

\section{Introduction}
Several classes of solar system bodies, comets \citep{Lisse1996}, planets \citep{Cravens2000,Dennerl2002a,Fujimoto2007,Snowden2009}, and even the interplanetary medium \citep{Cox1998,Snowden2004,Koutroumpa2007} are known sources of X-ray radiation. The X-ray emission common to all those objects is produced from the radiative de-excitation of highly-charged solar wind (SW) ions that have captured an electron from neutral particles encountered in these environments, and is accordingly named solar wind charge exchange (SWCX) X-ray emission. A thorough review of SWCX emission in the solar system is given in \cite{Bhardwaj2010} and \cite{Dennerl2010}.

In this paper, we focus on CX-induced X-ray emission from nonmagnetized planets, specifically from Mars. Martian CX emission was predicted for the first time by \cite{Cravens2000}, and the first estimates of the Martian SWCX emission based on comet observations and calculations were given by \cite{Krasnopolsky2001}. Following that, the first model of minor ion propagation through a simplified SW - Martian exosphere interface was devised by \cite{Holmstrom2001}. Thanks to the absence of an internal magnetic field, the SW can interact directly with the upper martian atmosphere, and CX collisions between multiply charged heavy SW ions and atmospheric neutral constituents may occur at high altitudes. The first detection of Martian X-rays was confirmed very soon by Chandra observations \citep{Dennerl2002a}. In addition to a fully illuminated disk of the size of the planet, a fainter halo was detected up to three Martian radii. Fluorescence and scattering of solar X rays are efficient processes at low altitudes, and they explain emissions from the disk, while X-ray inducing CX collisions provide a good description of the halo emission.

\cite{Gunell2004} estimated the CX X-ray emission from Mars in the context of the Chandra observations, incorporating a hybrid model for the SW-Mars interaction and a test particle simulation of heavy ion trajectories near Mars. They simplified the computation by considering a unique cross section for both H and O atoms, and replacing cascading transitions to ground state by a simplified two-step cascade model. This model assumes equal probabilities for all radiative transitions starting from the same initial excited state. The simulation shows good agreement with the Chandra 2001 observation on the total X-ray intensity. Chandra's limited spectral resolution, however, did not allow any conclusive spectral analysis.

Mars was also observed with the Reflection Grating Spectrometer (RGS) aboard XMM-Newton \citep{Dennerl2006}. RGS' spectral resolution allowed for the first time a detailed spectroscopic study of Martian X-rays from the disk and halo, resolving fluorescence spectral lines induced in the upper atmosphere from SWCX emission lines produced in collisions between SW heavy ions and exospheric neutrals at greater distances from the planet. This analysis yielded the strongest SWCX emission levels from Mars ever observed, with emission being detected up to eight Mars radii, and a total halo luminosity of (12.8$\pm$1.4) MW (mainly produced by SWCX emission lines).

More recently, \cite{Ishikawa2011} have observed Mars with the Suzaku XIS instrument in 2008, during the minimum phase of solar activity. However, their results did not show any significant SWCX emission from the planet, and only yielded a 2$\sigma$ upper limit of 4.3$\times$10$^{-5}$ photons cm$^{-2}$ s$^{-1}$ from the oxygen lines in the 0.5-0.65 keV energy range. In the same energy range, the Chandra and XMM-Newton observations yielded a solid 1$\sigma$ detection of 5.8$\times$10$^{-6}$ photons cm$^{-2}$ s$^{-1}$ and 3.6$\times$10$^{-5}$ photons cm$^{-2}$ s$^{-1}$ respectively. 

We present an improved model of the SWCX emission in Mars' environment, with an unprecedent spatial resolution (130 km) and detailed spectral information on the SWCX induced transitions that allows a direct comparison of the simulation results to the XMM-Newton measured photon fluxes and total luminosities. In section \ref{model} we describe the simulation model and initial solar and SW parameters appropriate for the XMM observation period derived from near-Earth instruments and extrapolated to Mars' orbit. In section \ref{results} we present projected emission maps, spectral line intensities, and total luminosities resulting from the simulations. Finally, in section \ref{discuss} we discuss how the model results compare to the data and previous simulations and offer some explanations for the discrepancies we found.


\section{Simulation model}\label{model}
The simulation process is separated into two main steps. In the first we use a three-dimensional (3D) and multi species hybrid model to provide a full description of the SW interaction with the Martian neutral environment. This simulation model has been designed by \cite{Modolo2005,Modolo2006}; the hybrid kernel of the simulation program makes use of the Current Advance Method and Cyclic Leapfrog scheme (CAM-CL) designed by \cite{Matthews1994}. We resolve the Maxwell equations and self-consistently obtain a quasi-stationary solution to the global electromagnetic (EM) field as well as a description of the plasma dynamic in the vicinity of the planet. This simulation model successfully reproduces the plasma environment surrounding Mars, such as the Martian bow shock (hereafter MBS) and the Magnetic pile-up boundary (hereafter MPB) observed by the Mars Global Surveyor and Phobos-2 spacecrafts \citep{Trotignon2006}. 

In a second step, once we achieve equilibrium in the hybrid simulation, the electric and magnetic field 3D distributions obtained with the hybrid model are used as the input field for a test-particle model. Trajectories of highly charged SW ions are followed in the simulation box, and X-ray emissions due to CX collisions with the planetary neutrals are computed. A similar procedure has been successfully used to investigate the capture of alpha particles by the Martian atmosphere \citep{Chanteur2009}.

\subsection{The hybrid model}
The hybrid model has been extensively described in \cite{Modolo2005,Modolo2006} and \cite{Kallio2011}. Solar and planetary ions are represented by sets of macroparticles, describing the full dynamics of each ion species, while electrons are described by a massless fluid that ensures the conservation of the charge neutrality of the plasma and contributes to the electronic pressure and the electric currents. The magnetic field undergoes a temporal evolution according to Faraday's law.

The coordinate system is defined such that the X axis points away from the Sun, (X = V$_{sw}/\left\|V_{sw}\right\|$) is aligned with the local SW direction, where $V_{sw}$ is the SW velocity. The Y axis lies in the ecliptic, perpendicular to $V_{sw}$ and pointing ahead on Mars' orbit. The interplanetary magnetic field (IMF) vector lies in the XY (ecliptic) plane, following Parker's spiral ($\phi$ = -56$^{\circ}$ with respect to X axis at Mars). Finally the Z axis is toward the motional electric field, defined by E$_{conv}$ = -$V_{sw} \times B_{IMF}$, and therefore coincides with the north ecliptic pole.

The spatial resolution of the 3D uniform simulation grid is 130 km, equal to the inertial length of the SW protons. The planet is modeled as a fully absorbing obstacle with a radius of 3400 km, and ions penetrating in the obstacle boundary are stopped. Open boundaries are used in the SW direction and periodic boundaries in the perpendicular directions, except for the planetary ions, which are escaping freely from the simulation domain. At the first step of the simulation a computation is run from time t = 0 to time t = 1000 s, while a nearly stationary solution is obtained around time t = 450 s.

This hybrid simulation is performed for SW and EUV conditions corresponding to the period of the XMM-Newton observation of Mars starting on Nov. 19, 2003 23:50 UT. The different parameters are reviewed below and summarized. The SW plasma parameters used for this simulation are provided by the Advanced Composition Explorer (ACE) and Wind experiment and corrected to the Martian orbit. 

\subsubsection{Solar parameters}\label{solar}
\begin{figure}
   \centering
   \includegraphics[width=0.8\linewidth]{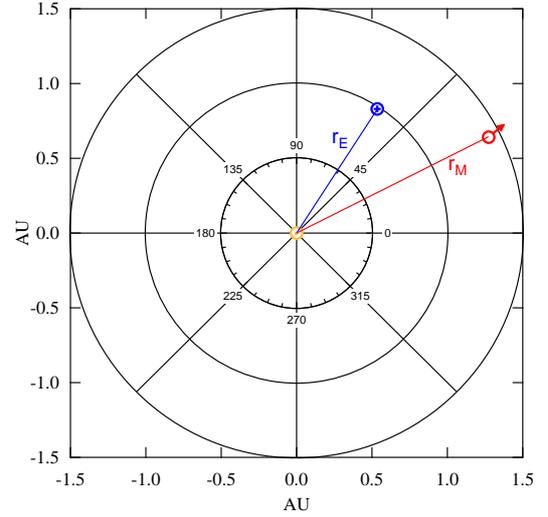}
      \caption{Relative position of Earth and Mars on November 20, 2003, generated by the IMCCE (http://www.imcce.fr/en/ephemerides/). A color version of this figure is available in the online journal.)
              }
         \label{FigGeo}
   \end{figure}

The ACE and Wind spacecrafts are located at the Sun-Earth's L1 point, and Figure~\ref{FigGeo} presents the relative position for Mars, Earth, and Sun at the time of the XMM observation. At first order, Mars and the Earth are in the same sector, meaning that near-Earth solar and SW monitoring can be used as a proxy to determine the SW and EUV conditions near the planet.

\begin{figure}
   \centering
   \includegraphics[width=\linewidth]{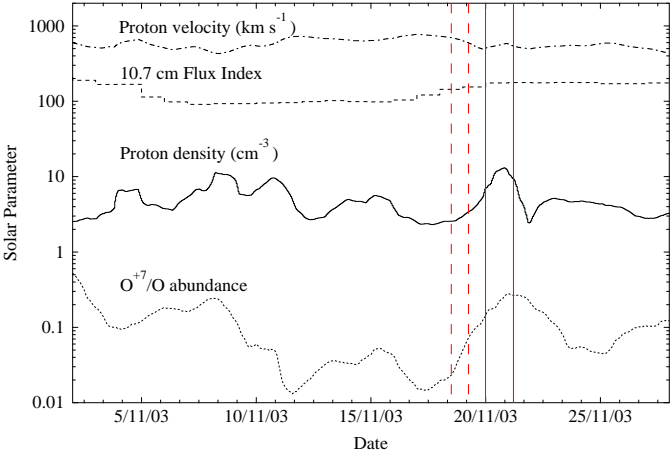}
      \caption{Solar parameters measured by near-Earth instruments during the three months preceding the XMM-Newton observation of Mars in November 2003. From top to bottom the curves represent, averaged over 3 solar rotations (81 days): proton velocity in km s$^{-1}$ (dot-dashed), the F10.7-cm flux index (dashed), proton density in cm$^{-3}$ (solid) and O$^{+7}$/O relative abundance (dotted). In-situ data are from ACE at L1, while the radio flux indices were taken from the NOAA Space Weather Prediction Center. The solid vertical lines show the start and end of the XMM-Newton observation of Mars, and the dashed vertical lines show the SW event window applied to the simulations, once the Earth-Mars SW delay time is taken into account.}
         \label{FigSW}
   \end{figure}

In Figure~\ref{FigSW} we present the F10.7-cm Radio flux index from the NOAA Space Weather Prediction Center\footnote{http://www.swpc.noaa.gov/Data/index.html} and SW in-situ data from ACE Science Center Level 2 database\footnote{http://www.srl.caltech.edu/ACE/ASC/level2/} for the period of November 2003, when the XMM-Newton observation of Mars took place. If we ignore the longitudinal difference in Mars' position compared to Earth (ACE) and assume a constant SW propagation velocity between the L1 point and Mars, then the time delay that needs to be applied to the ACE in-situ data in order to find the appropriate SW event window at Mars is $\Delta t = \Delta r/V_{sw}$, where $\Delta r = r_M - r_E$ is the heliocentric distance difference between Mars and the L1 point. We find that this corresponds approximately to the window between 2003/11/18 12:00~UT and 2003/11/19 06:00~UT (Figure~\ref{FigSW}). We derive the SW parameters by averaging the ACE data over this window and scaling to Mars' heliocentric distance for the particle densities ($1/r^2$).

The main ion species in the SW is H$^+$ and simulation values have been normalized with respect to H$^+$ parameters. The main SW plasma parameters adjusted to the Martian orbit in the event window defined above are
\[
\begin{array}{lp{0.8\linewidth}}
Density: & $n_{sw} \sim$ 2 cm$^{-3}$\\
Bulk\, speed: & $V_{sw} \sim$ 675 km s$^{-1}$\\
Thermal\, speed: & $v_{th} \sim$ 80 km s$^{-1}$\\
Electron\, temperature: & T$_e$ = 3.\, 10$^5$ K\\
He^{++}\, abundance & 4$\%$\\
IMF: & $\left\|B_{IMF}\right\|$ = 3 nT,\, $\phi$ = -56$^{\circ}$\\
\end{array}
\]

The F10.7-cm index can be used as proxy for the EUV solar conditions that influence the planet's neutral environment. During the November 20, 2003 period the F10.7-cm index is 175 and the 81 day (three solar rotations) average is 130, suggesting that the conditions are close to solar maximum. The neutral environment is then chosen adequately.
 
\subsubsection{Neutral atmosphere and exosphere}
\begin{figure}
   \centering
   \includegraphics[width=\linewidth]{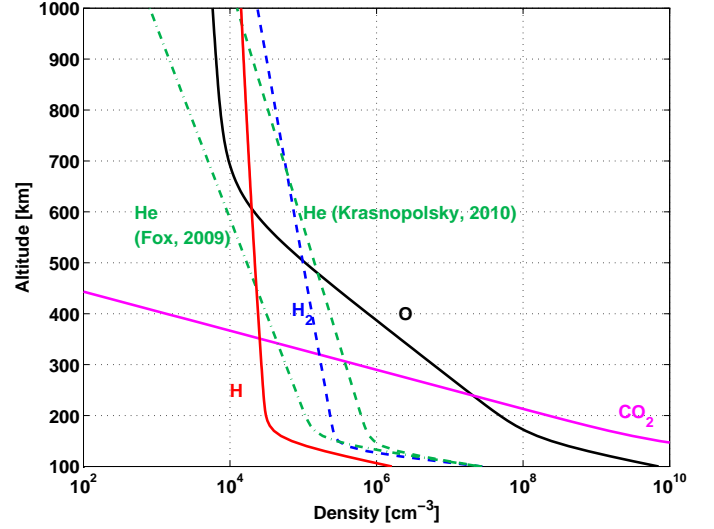}
      \caption{Neutral density profiles for the CO$_2$ (magenta), H (red), and O (black) coronas used in our simulations. The H$_2$ (blue dashed) and He (green dashed) coronas from \cite{Krasnopolsky2010} and He (green dash-dotted) corona from \cite{Fox2009} are also plotted for comparison. (A color version of this figure is available in the online journal.)
              }
         \label{FigNcor}
   \end{figure}

The Martian neutral environment is described by three coronae of CO$_2$, H, and O show in Figure~\ref{FigNcor}. Both H$_2$ and He coronas are also thought to be important at altitudes from around 400 km to 1000 km \citep[also presented in Figure~\ref{FigNcor}, extracted from][]{Fox2009, Krasnopolsky2002, Krasnopolsky2010}. However, the density profiles of these neutral populations decrease very quickly with altitude and should only have an effect in the X-ray emission in the disk region, as we discuss in Section~\ref{discuss}. This version of the model does not include the H$_2$ and He coronas, but future work will include these components.

All information is based on input from the ``Solar Wind Model Challenge" international team at the International Space Science Institute (ISSI) in Bern \citep{Brain2010}. Atmospheric and exospheric coronas are partly deduced from the Mars thermosphere global circulation model \citep[MTGCM,][]{Bougher2000,Bougher2006,Bougher2008}, unless noted otherwise, assuming a spherically symmetric neutral environment. The MTGCM was run for Martian longitude position Ls = 270 and F10.7 = 105. Although observed EUV conditions are slightly more intense than those used for the atmospheric simulations, these simulations are the most approriate runs.

For CO$_2$, the density profile used is
   \begin{equation}
      n_{CO_2} = 5.88\,10^{18}\,exp\left(-\frac{z}{7.00}\right)\, +\, 3.55\,10^{13}\,\exp\left(-\frac{z}{16.67}\right), \\
   \end{equation}
which is in very good agreement with the models of \cite{Krasnopolsky2002} and \cite{Fox2009}. In the equation, z is the altitude in km and n is the density in cm$^{-3}$. 

For O, the `cold' oxygen density profile used represents a fit from the MTGCM results:
   \begin{equation}
      n_{O}^{cold} = 2.33\,10^{13}\,exp\left(-\frac{z}{12.27}\right)\, +\, 2.84\,10^{9}\,\exp\left(-\frac{z}{48.57}\right). \\
   \end{equation}
Below 400~km the computed density profile for thermal O atoms agrees well with results of the photo chemical model by \cite{Krasnopolsky2010}. The `hot' density profile used is a fit from \cite{Valeille2010} results:
   \begin{eqnarray}
      n_{O}^{hot} & = & 1.56\, 10^{4}\,exp\left(-\frac{z}{696.9}\right)\, +\, 2.92\,10^{3}\,\exp\left(-\frac{z}{2891.}\right)\, \nonumber \\
      						&		&	+\, 5.01\,10^{4}\,\exp\left(-\frac{z}{99.19}\right).
   \end{eqnarray}

For hydrogen, the density profile used is a combination of results from \cite{Anderson1971} and \cite{Krasnopolsky2002}:
   \begin{eqnarray}
      n_{H} & = & 10^{3}\,exp\left[9.25\, 10^5\, \left(\frac{1}{z\, +\, 3393.5}\, -\, \frac{1}{3593.5}\right)\right] \, \nonumber \\
      & & +3\, 10^{4}\,exp\left[1.48\, 10^4\, \left(\frac{1}{z\, +\, 3393.5}\, -\, \frac{1}{3593.5}\right)\right].
   \end{eqnarray}

\subsection{Test particle simulation}
The second step consists of running a test particle simulation for each minor species of the SW with the electric and magnetic fields obtained from the hybrid simulation model. We consider that the E and B fields are in equilibrium so we no longer introduce H$^+$ and He$^{++}$ ions in the simulation box, but only minor SW ions (C, N, O). In the test particle simulation for a given minor species of the SW, we follow the trajectories of N$=2\times10^6$ macroparticles having the same initial statistical weight when injected into the simulation box. This initial statistical weight is computed by considering that the physical flux corresponding to the N macroparticles injected during one second should be equal to the nominal flux of the given minor species in the SW. At each time step, meaning at each position of the particle, CX collisions are computed in the cell, and the statistical weight of the test-particle is reduced to mimic the loss of the parent ions caused by collisions. We inject one heavy SW ion macroparticle after the other in the box and solve equations of motion to calculate its trajectory. The next macroparticle is injected when the former one has either left the box, is absorbed by the planet/obstacle, or its weight is reduced to 0 because of successive CX collisions. This technique has been successfully applied to investigating the helium budget in the Martian atmosphere \citep{Chanteur2009}.The reaction the SW ion X$^{+q}$ undergoes in the simulation is
\begin{equation}\label{eq6}
X^{+q}\, +\, M \rightarrow\, X^{+(q-1)*}\, +\, M^+ \\
\end{equation}
where M = [H, O] are the major neutral target atoms in the Martian environment. In each grid cell we calculate the probability of ion X$^{+q}$ capturing an electron from exospheric neutrals, register the ion X$^{+(q-1)*}$ production rate, and cumulate the contributions of ions X$^{+(q-1)*}$ produced by all N X$^{+q}$ ions in the cell. 

As long as it may collide with exospheric neutrals, i.e. in practice as long as it remains within the simulation box around Mars, the X$^{+(q-1)}$ ion produced in reaction \ref{eq6} can continue to exchange charges until it is neutralized. We investigated the influence of these secondary CX collisions  for the case of O$^{+6}$ ions sequentially produced by O$^{+8}$ and O$^{+7}$ ions in Mars' exosphere. The SWCX emission produced by secondary CX collisions has been estimated and was found to contribute to the total O$^{+6}$ (OVII) emission by less than 1\%, hence can be ignored.

For all minor SW species the initial distribution function in the SW is an isotropic Maxwellian with a bulk velocity of 675 km/s. Although the SW minor ions do not influence the EM structure equilibrium around Mars, the SWCX line emission produced by the reactions described above is directly proportional to the parent ion flux. More precisely, the SWCX line photon flux is the integral over the line of sight (here limited to the simulation box size) of the product of the minor ion density, the neutral density (here the exospheric H and O), the relative velocity between the two colliding particles (usually averaged to the SW ion velocities), the CX collision cross-section, and the individual line emission probability \citep[based on theoretical calculations by][]{Kharchenko2000}. 

In Table~\ref{table:1} we summarize the parameters for the minor SW ions included in the simulations. The first column lists the ion charge state. Columns 2 and 3 list the charge state relative abundances with respect to oxygen, and alpha (He$^{++}$) to proton ratio for the typical slow SW \citep{Schwadron2000} and the real-time values extracted from the ACE database for the window of the XMM observation of Mars, as described in section \ref{solar}. The ion density N$_i$ is then defined as the product:  

\begin{equation}
N_i = \left[\frac{X^{+q}}{O}\right]\, \left[\frac{O}{He}\right]\, \left[\frac{He}{H}\right]\, n_{sw}\,  .
\end{equation}

In columns 4 and 5 of Table~\ref{table:1} we also summarize the cross sections used in the simulations for CX reactions with H and O atoms, respectively. The cross section values for H are assembled from references listed in Table~1 of \cite{Koutroumpa2006}. For the cross sections in CX with O atoms, we adopt the values listed in \cite{Schwadron2000} for reactions with H$_2$O molecules. All cross sections refer to typical slow SW velocities of 450~km~s$^{-1}$. The CX cross sections are known to vary smoothly with the collision velocity, according to theoretical \citep[e.g.,][]{Harel1998,Lee2004} and laboratory \citep[e.g.,][]{Beiersdorfer2001,Mawhorter2007} results. Nevertheless, the precise dependence on the collision velocity is still rather uncertain for a number of ions, and in general it is not expected to vary significantly for speeds above 400~km~s$^{-1}$, according to the same works. Moreover, preliminary tests with velocity-dependent cross sections revealed negligible differences in our simulation results, in accordance with the \cite{Gunell2004} conclusions on the matter as well.

\begin{table}
\caption{Parameters of SW minor ions included in simulations}             
\label{table:1}      
\centering                          
\begin{tabular}{l c c c c}        
\hline\hline                 
            Ion &  \multicolumn{2}{c} {[X$^{+q}$/O] Abundance} &  \multicolumn{2}{c}{Cross section (10$^{-15}$ cm$^{2}$)} \\ 		
             & Slow SW\tablefootmark{a} & Real-Time\tablefootmark{b} &  CX with H & CX with O\tablefootmark{c} \\ 		
\hline                        
            C$^{+6}$ &  0.318  & 0.13$\pm$0.06 & 4.2 & 5 \\
            C$^{+5}$ &  0.21    & 0.37$\pm$0.03 &  4.0 & 2 \\
            N$^{+7}$ &  0.006  &   no data       &  5.7 & 12 \\
            N$^{+6}$ &  0.058  &   no data       &  3.7 & 5 \\
            O$^{+8}$ &  0.07    & $\le$0.002 & 5.7 & 6 \\
            O$^{+7}$ &  0.20    & 0.04$\pm$0.02 & 3.4 & 12 \\
            He$^{++}$ &  78$\pm$4  & 92$\pm$3 &  &  \\
\hline                                   
            He/H &  0.044$\pm$0.003  & 0.045 &  &  \\

\hline                                   
\end{tabular}
\tablefoot{ 
\tablefoottext{a}{From \cite{Schwadron2000}}
\tablefoottext{b}{From ACE-SWICS}
\tablefoottext{c}{From \cite{Schwadron2000}, based on the assumption that atomic O has similar weight to H$_{2}$O.}
}
\end{table}

\section{Simulation results}\label{results}
We calculated CX-induced X-ray emission maps in the same geometry conditions as the XMM-Newton 2003 observation for the most abundant SW ions (listed in Table~\ref{table:1}), in particular the ones whose emission lines were identified in the XMM data. On this date, Mars was at a helioecliptic longitude of about 27$^{\circ}$ and the Earth (XMM) at a longitude of 57$^{\circ}$, at a phase angle (Sun-Mars-Earth) of $\sim$40$^{\circ}$.

\subsection{Emission morphology}
\begin{figure}
   \centering
   \includegraphics[width=0.8\linewidth]{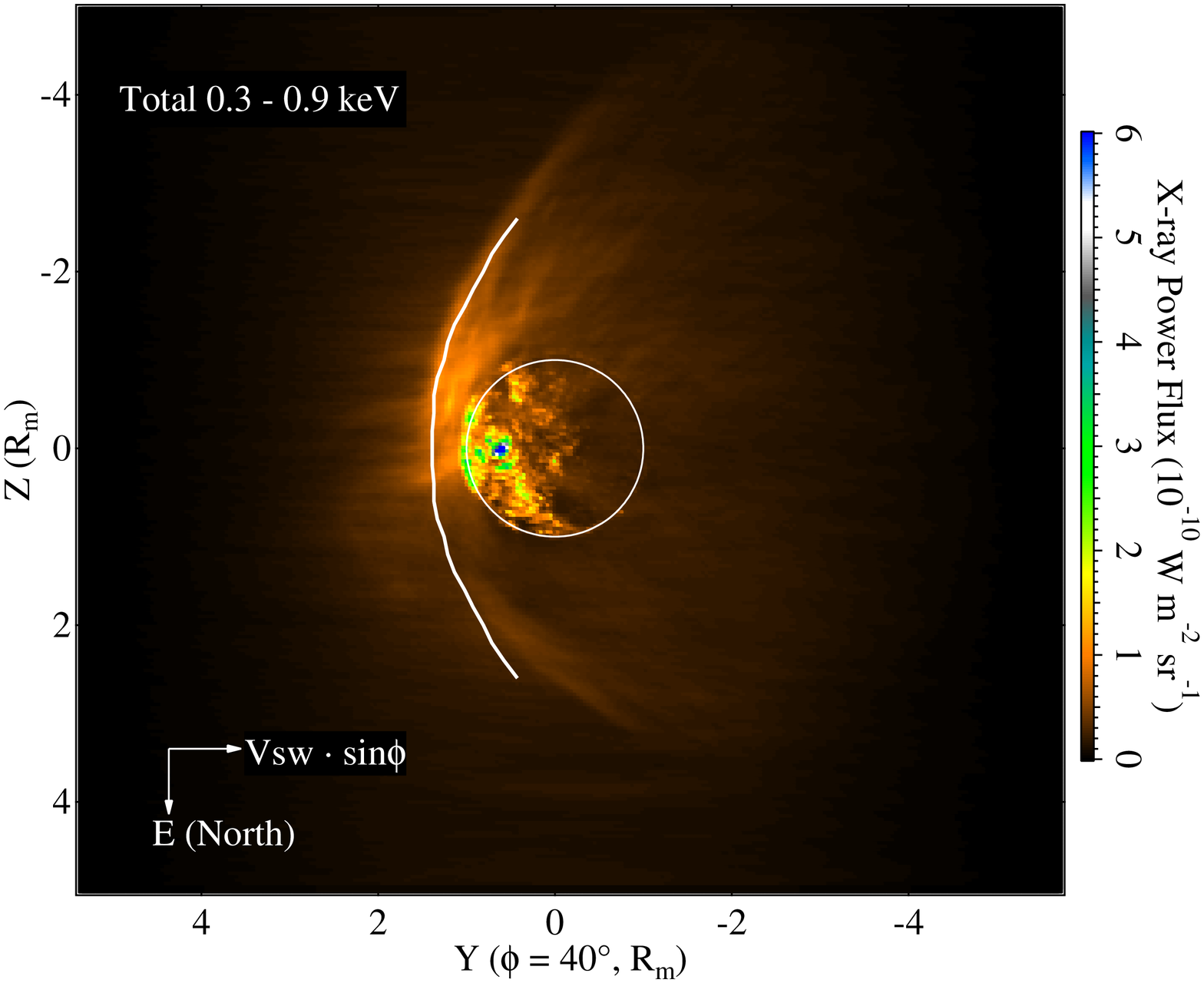}
   \includegraphics[width=0.8\linewidth]{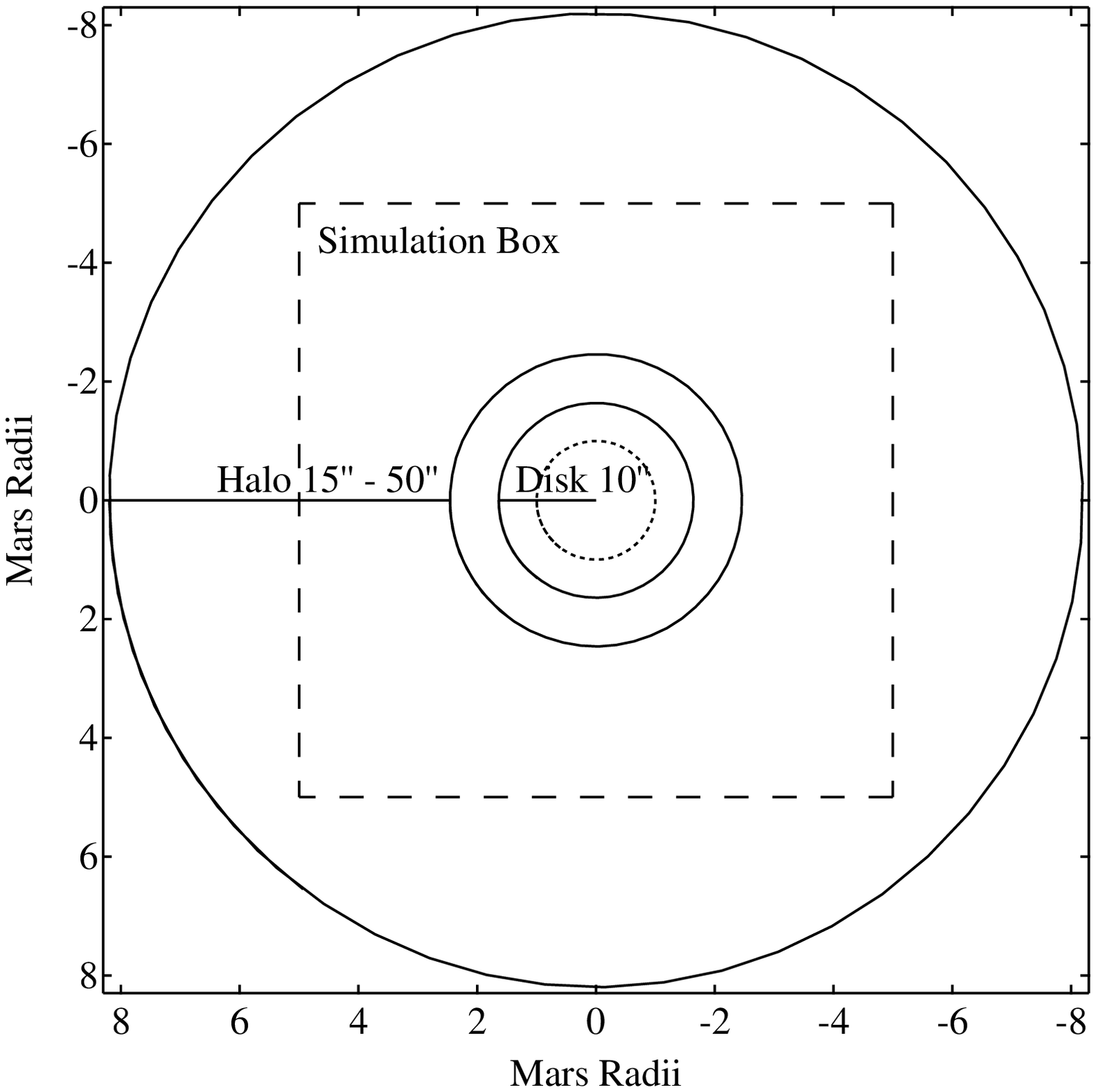}
   \includegraphics[width=0.8\linewidth]{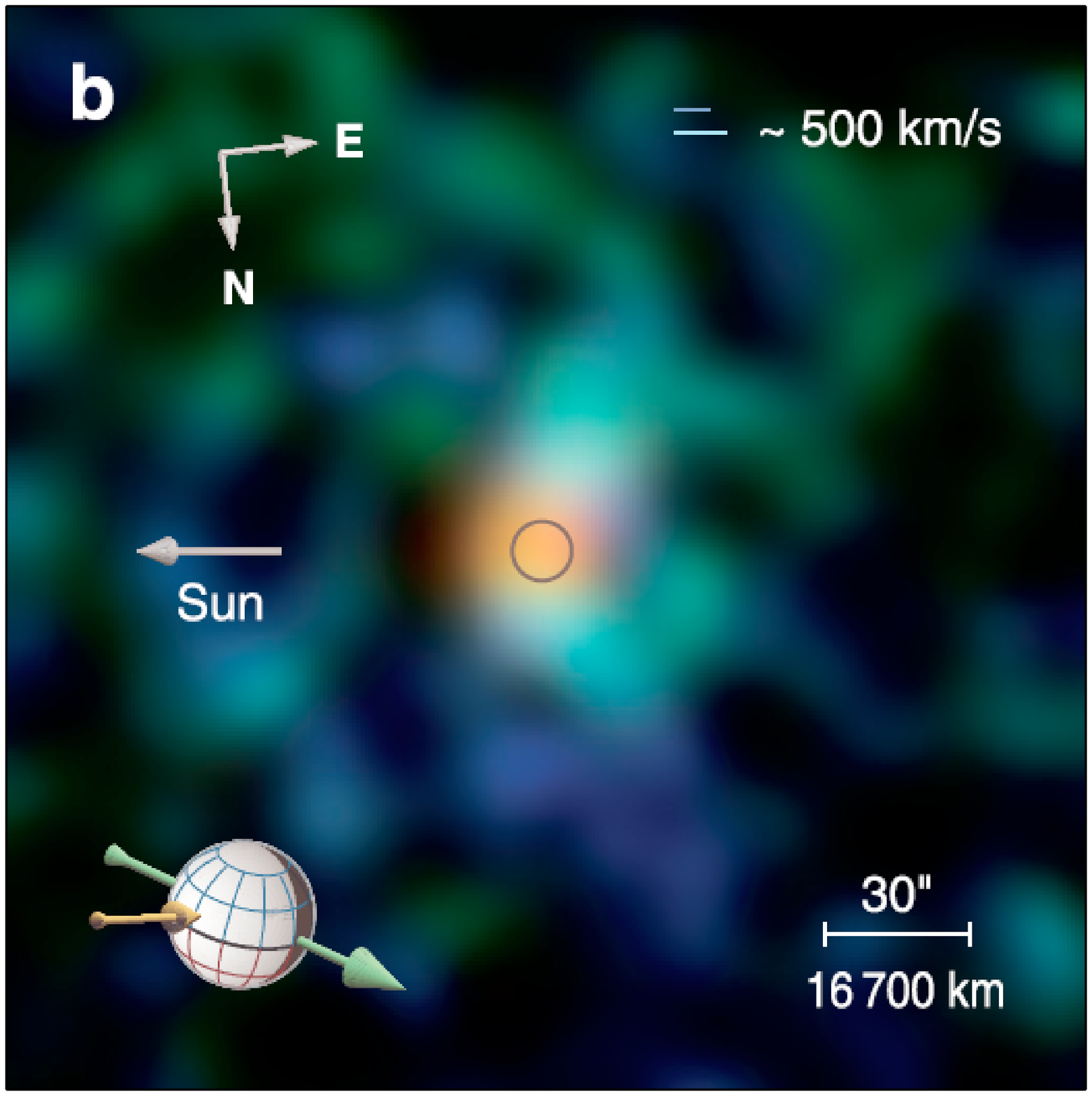}
      \caption{\textit{Top - }Total CX-induced X-ray power flux map from Mars in units of W~m$^{-2}$~sr$^{-1}$, in the same projection as the XMM-Newton observation, at a $\phi = 40^{\circ}$ phase angle. Axes are in Mars radii. The planet is represented by the white circle, and the white curve marks the MBS position based on MGS data \citep{Vignes2000}. The B-Vsw plane corresponds to the ecliptic and the electric field (E) axis points north. North points downward for direct comparison with \cite{Dennerl2006}'s figure 8 reproduced in the bottom panel.
      \textit{Middle - }Size of our simulation box projection at a 40$^{\circ}$ angle (dashed square), compared to Mars' X-ray disk ($r<10''$) and halo ($15''< r < 50''$) sizes defined by \cite{Dennerl2006}. Mars is represented by the dotted circle.
     \textit{Bottom - }Reproduction of \cite{Dennerl2006}'s figure 8. The scale is not the same as in the top two panels. (A color version of this figure is available in the online journal.)
              }
         \label{FigAllMap}
   \end{figure}

In the top panel of Figure~\ref{FigAllMap} we present the projected map of the total power flux (W~m$^{-2}$~sr$^{-1}$) for the spectral lines included between 0.3 and 0.9~keV. The white curve superimposed on the map represents the average position of the MBS  in the same projection as the XMM observations as inferred from the Mars Global Surveyor measurements \citep{Vignes2000}. In the middle panel of Figure~\ref{FigAllMap} we compare the size of our simulation box to the field-of-view of RGS that was used to extract X-ray spectra from Mars. In the bottom panel we reproduce figure 8 from \cite{Dennerl2006}, showing the total (fluorescence in orange and SWCX in green and blue) emission from Mars, for comparison. 

The model emission displays the characteristic crescent-like structure on the sunward side of the planet, where the plasma discontinuity at the MBS and the flow of SW around the planet-obstacle are clearly identified. Indeed, as shown in the top panel of Figure~\ref{FigAllMap}, the discontinuity in the predicted SWCX emission in the model is consistent with the average position of the MBS as measured by the MGS \citep{Vignes2000}. However, while the MGS-inferred MBS is axisymmetric around the Sun-Mars axis, the crescent-like structure of the modeled emission is asymmetric, owing to the influence of the motional electric field. The MBS boundary in the simulations (as illuminated by the SWCX emission) is pushed towards the Sun in the southern hemisphere, and pushed towards the planet near its north pole. More details on the plasma boundaries and the output of the hybrid code can be found in \cite{Modolo2005,Modolo2006}.

The overall shape of the model emission generally looks similar to the observed total emission pattern \citep[][]{Dennerl2006}. The brightest part of the emission is observed by projection on the planet's disk (again, Mars is observed at a 40$^{\circ}$ phase angle), and seems to be produced very close to the planet's surface at the limb in the sunward direction. The fainter halo is observed up to 4 Mars radii in the north-south direction, while \cite{Dennerl2006} claim a much larger extend, up to 8 Mars radii. Due to the asymmetry of the MBS, the emission pattern is brighter and more elongated toward the south pole of the planet, along the direction opposite the electric field vector, while there seems to be a slight `depression' of the emission towards the northern hemisphere of the planet.

All the individual ion emission maps (not presented here) show an equivalent emission pattern. Although the strong asymmetry predicted by the model could have been perceived as a lobe-like structure by RGS considering the detector's spatial resolution \citep[similar to the oxygen line maps presented in Figure 7 of][]{Dennerl2006}, overall the emission pattern seems rather continuous in the simulations. The qualitative agreement between the emission pattern in the simulations, and the data seems to support the idea that the observed tilt of the upper and lower wings in the data indeed has a morphological nature \citep[see discussion in Section~5.1 of ][]{Dennerl2006}.

\subsection{Luminosity and spectral analysis}
The list of lines and their photon fluxes calculated with the model are presented in Table~\ref{table:2}. In column 1 we list the emitting ion, and in columns 2 and 3 the identified transition and energy (in eV) respectively. Columns 4, 5, and 6 list the simulated line fluxes for the complete simulation box, disk, and halo. We adopt the same definition as \cite{Dennerl2006} for the disk (r $\leq 10''$) and the lower altitude limit for the halo ($15'' \leq$ r), while the upper limit for our halo is defined by the simulation box limits (see Figure~\ref{FigAllMap}-bottom). Here we compare only CX emission lines; i.e., fluorescence lines are not included in the analysis. Model photon fluxes are computed separately for the disk and halo components, as well as for the total box size and given in units of 10$^{-6}$~cm$^{-2}$~s$^{-1}$. Finally columns 7, 8, and 9 list the fitted photon fluxes in the CX identified lines from the XMM-Newton observation \citep[extracted from Table 2 in][]{Dennerl2006} for the total, disk, and halo regions. 

Obviously, although the model includes all the major transitions detected by XMM, it widely underestimates the data line fluxes, which are in general more than one order of magnitude higher than the model line fluxes. For energies between 0.3 and 0.9~keV, the total luminosity within the simulation box is 0.23~MW, while for the disk we find  0.06~MW and for the halo 0.12~MW.

\begin{table*}
\caption{SWCX emission lines in the simulations and RGS spectra of Mars and its halo}             
\label{table:2}      
\centering                          
\begin{tabular}{l l c c c c c c c}        
\hline\hline                 
            Ion &  Transition\tablefootmark{a}  &Energy (eV)  &  \multicolumn{6}{c}{Line Photon flux (10$^{-6}$~cm$^{-2}$~s$^{-1}$)} \\ 		
           &  &  &\multicolumn{3}{c}{Model} &  \multicolumn{3}{c}{Data\tablefootmark{b}}\\ 		
             &  &  &Total & Disk & Halo & Total & Disk & Halo \\ 		
\hline                        
NeVIII	&2p $\rightarrow$1s\tablefootmark{c}	&872.5	&		&		&		&(1.8$\pm$1.8)		&(0.5$\pm$0.5)		&3.8$\pm$2.2\\		
OVIII		&7p $\rightarrow$1s					&853	&4.2e-6	&1.0e-6	&2.3e-6	&				&				&	\\					
OVIII		&6p $\rightarrow$1s					&846.6	&5.6e-5	&1.4e-4	&3.1e-4	&				&				&	\\					
OVIII		&5p $\rightarrow$1s					&835.9	&1.9e-3	&4.7e-4	&1.1e-3	&				&				&	\\					
OVIII		&4p $\rightarrow$1s					&816.3	&6.3e-4	&1.5e-4	&3.5e-4	&3.9$\pm$1.5		&(0.4$\pm$0.4)		&(1.6$\pm$1.6)\\
OVIII		&3p $\rightarrow$1s					&774	&1.7e-3	&4.2e-4	&9.6e-4	&				&				&	\\					
OVII		&6p $\rightarrow$1s					&720.9	&9.0e-5	&2.8e-5	&4.3e-5	&				&				&	\\					
OVII		&5p $\rightarrow$1s					&712.3	&2.3e-5	&7.0e-6	&1.1e-5	&7.9$\pm$2.1		&(0.4$\pm$0.4)		&6.2$\pm$1.7\\
OVII		&4p $\rightarrow$1s					&697.8	&0.01	&4.3e-3	&6.8e-3	&				&				&	\\					
OVII		&4p (T) $\rightarrow$1s				&697	&1.3e-4	&4.0e-5	&6.3e-5	&				&				&	\\					
OVII		&3p $\rightarrow$1s					&665.6	&0.02	&5.9e-3	&9.3e-3	&				&				&	\\					
OVII		&3p (T) $\rightarrow$1s				&663.9	&2.4e-4	&7.5e-5	&1.2e-4	&				&				&	\\					
OVIII		&2p $\rightarrow$1s					&653.1	&0.01	&3.3e-3	&7.5e-3	&9.9$\pm$2.4		&4.3$\pm$2.1		&7.4$\pm$2.1\\
NVII		&6p $\rightarrow$1s					&648.2	&4.9e-4	&1.3e-4	&2.6e-4	&				&				&	\\					
NVII		&5p $\rightarrow$1s					&640	&6.0e-3	&1.5e-3	&3.2e-3	&				&				&	\\					
NVII		&4p $\rightarrow$1s					&625	&3.0e-3	&7.7e-4	&1.6e-3	&				&				&	\\					
NVII		&3p $\rightarrow$1s					&592.6	&5.2e-3	&1.3e-3	&2.8e-3	&				&				&	\\					
OVII-r	&2p $\rightarrow$1s					&574	&0.03	&7.5e-3	&0.01	&(2.3$\pm$2.3)		&7$\pm$3.3		&(1.4$\pm$1.4)\\
OVII-i	&2p (T) $\rightarrow$1s				&568.5	&0.03	&0.01	&0.02	&1.5$\pm$1.5		&(1.2$\pm$1.2)		&2.2$\pm$2.2\\
OVII-f	&2p (T) $\rightarrow$1s				&560.9	&0.21	&0.06	&0.10	&12.8$\pm$3.2		&4.2$\pm$2.9		&5.3$\pm$2.3\\
NVI		&4p $\rightarrow$1s					&521.6	&0.02	&5.6e-3	&0.01	&				&				&	\\					
NVI		&4p (T) $\rightarrow$1s				&520.9	&8.6e-5	&2.4e-5	&4.5e-5	&				&				&	\\					
NVII		&2p $\rightarrow$1s					&500	&0.05	&0.01	&0.02	&9.3$\pm$2.7		&4.5$\pm$2.2		&7.1$\pm$2.2\\
NVI		&3p $\rightarrow$1s					&497.9	&0.01	&3.9e-3	&7.2e-3	&				&				&	\\					
NVI		&3p (T) $\rightarrow$1s				&496.5	&7.8e-5	&2.2e-5	&4.1e-5	&				&				&	\\					
CVI		&6p $\rightarrow$1s					&476.2	&3.7e-4	&9.5e-5	&2.0e-4	&				&				&	\\					
CVI		&5p $\rightarrow$1s					&470.2	&0.03	&8.4e-3	&0.02	&5.5$\pm$2.2		&2.7$\pm$2.3		&5.5$\pm$1.8\\
CVI		&4p $\rightarrow$1s					&459.2	&0.16	&0.04	&0.08	&4.5$\pm$3.2		&5.7$\pm$2.8		&5.3$\pm$2.5\\
CVI		&3p $\rightarrow$1s					&435.4	&0.10	&0.03	&0.05	&6.6$\pm$2.3		&(1.9$\pm$1.9)		&7.9$\pm$2.7\\
NVI		&2p $\rightarrow$1s					&430.7	&0.04	&0.01	&0.02	&				&				&	\\				
NVI		&2p (T) $\rightarrow$1s				&426.1	&0.04	&0.01	&0.02	&				&				&	\\					
NVI		&2s (T) $\rightarrow$1s				&419.8	&0.27	&0.08	&0.14	&				&				&	\\					
CV		&5p $\rightarrow$1s					&378.5	&		&		&		&9.9$\pm$4.2		&(2.8$\pm$2.8)		&8.4$\pm$3.1\\			
CV		&4p $\rightarrow$1s					&370.9	&0.03	&0.01	&0.02	&				&				&	\\					
CV		&4p (T) $\rightarrow$1s				&370.4	&5.5e-5	&1.8e-5	&2.8e-5	&				&				&	\\
CVI		&2p $\rightarrow$1s					&367.4	&0.61	&0.16	&0.33	&16.2$\pm$5.2		&11$\pm$4.2		&9.1$\pm$3.5\\
CV		&3p $\rightarrow$1s					&354.5	&0.08	&0.02	&0.04	&				&				&	\\					
CV		&3p (T) $\rightarrow$1s				&353.4	&1.8e-4	&5.7e-5	&9.0e-5	&				&				&	\\					
OVIII		&2s $\rightarrow$1s					&326.5	&1.4e-3	&3.3e-4	&7.6e-4	&				&				&	\\					
CV		&2p $\rightarrow$1s					&307.9	&0.16	&0.05	&0.08	&				&				&	\\					
CV		&2p (T) $\rightarrow$1s				&304.2	&0.07	&0.02	&0.04	&				&				&	\\					
  
\hline                                   
\multicolumn{3}{l}{Total luminosity (MW)}						&	0.23		&	0.06		&	0.12		&11.8\tablefootmark{c}&5.8\tablefootmark{c}&9.1\tablefootmark{d}	\\
\hline                                   
\end{tabular}
\tablefoot{ 
\tablefoottext{a}{Transitions involving a triplet state are noted with a (T).}
\tablefoottext{b}{Only the SWCX emission lines from \cite{Dennerl2006} are reported in this table.}
\tablefoottext{c}{This line has been falsely identified by \cite{Dennerl2006} as the NeVIII 2p $\rightarrow$1s transition because it would involve the electron being captured in an excited core level, which cannot be produced by CX.}
\tablefoottext{d}{Excluding the NVII 3p$\rightarrow$1s due to its large uncertainty.}
}
\end{table*}

\subsection{The OVII triplet}
The ratio G = (OVII-f +OVII-i)/OVII-r of triplet-to-singlet transitions of the He-like OVII ion is used to identify the emission mechanism and is usually less than one for hot plasmas where excitation of O$^{+6}$ ions occurs due to electron impact. This ratio is in general greater than three in the case of CX \citep{Kharchenko2003}. In the XMM 2003 observations of Mars, the ratio is found G$\sim$ 6 for the total emission and G $\sim$ 5 for the halo region. Our model reflects the atomic data we have used, and yields a mean ratio of $\sim$ 8 in the simulation box, which is consistent with the values found by \cite{Dennerl2006}. 

However, \cite{Dennerl2006} report a much fainter G ratio (0.8$\pm$0.6) for oxygen lines from the disk. This interesting result may be interpreted in two ways: either a different mechanism is responsible for the disk emission or the G factor differs from its theoretical nominal value for some physical reason. Because the forbidden line emission disappears, it suggests an effect of collisional depopulation. We did investigate this possibility.

If N$_m$ is the density of the ions in the metastable level, then the combined effects of production through CX collisions and depopulation by radiative decay and collisions imply that
   \begin{equation}
      \frac{dN_m}{dt} = \sigma _{cx}\, n\, V\, N_i\, -\, N_m \frac{1+\omega _{qc}\, \tau _r}{\tau _r}  
   \end{equation}
where $\sigma _{cx}$ is the CX cross-section, n the density of neutrals (H, O), V is the collision velocity, N$_i$ the ion density (O$^{+7}$ in this case), $\omega_{qc}$ represents the quenching collision frequency, which includes all types of collisions and all target atoms, electrons and ions able to induce the quenching, and $\tau _r$ is the lifetime of the metastable state against radiative decay.

Quenching collisions remove excited electrons from the long-lived metastable states and, because of that, suppress their emission. Degradation of the X-ray emission of metastable ions may be caused by subsequent CX collisions of metastable O$^{+6*}$ ions with atmospheric gas. These CX collisions lead to a formation of doubly excited ion states O$^{+5**}$ that decay mostly due to Auger process. If $\sigma _q$ is a cross-section representative of all collisions for all colliding particles i of total density n$_q$ in the planet's atmosphere, then,
   \begin{equation}
      \omega _{qc}  = \sum_{qi} n_{qi} \sigma _{qi}\, V = n_q \sigma _{q}\, V   \,.
   \end{equation}
For the steady state solution, dN$_m$\,/\, dt = 0,
   \begin{equation}
      N_m  = \frac{\sigma_{cx}\, n\, V\, N_i}{1\, +\, \omega_{qc} \tau_r} \tau_r  
   \end{equation}
and the line radiation flux per unit volume is
   \begin{equation}
      I  = \frac{I_o}{1\, +\, \omega_{qc} \tau_r}  
   \end{equation}
where I$_o$ is the intensity in absence of collisions.

Collisions become important and significantly reduce the radiation when $\omega _{qc}\, \tau _r $ = 1, i.e. n$_q$ = $(\sigma _{q}\, V\, \tau_r)^{-1}$. For the lifetime of the O$^{+6*}$ metastable state $\tau_r$ = 10$^{-3}$~s, ion velocities through the atmosphere of V = 400~km~s$^{-1}$, and assuming $\sigma_q$ = 10$^{-15}$~cm$^{2}$ based on a simple geometrical scaling and cross-section values of the CX collisions of O$^{+6}$ ions with the CO$_2$ and CO ($5.\,10^{-15}$~cm$^2$), the corresponding critical density is n$_q$ = $2.5\,10^{10}$~cm$^{-3}$. This critical  density is relevant to the altitude of 150~km where the atmospheric gas is mostly represented by CO$_2$ molecules. The same calculation for ions fully decelerated down to 4~km~s$^{-1}$ gives n$_q$ = $2.5\, 10^{12}$~cm$^{-3}$, which corresponds to altitudes on the order of 100km. The cross-section $\sigma_q$ slightly increases at low ion velocities, but it would not change the quenching altitude estimations dramatically because of the exponential character of the atmospheric density profiles.

As a conclusion, quenching of the 2$^3$S$_1$ OVII metastable state and degradation of the forbidden line OVII-f, i.e. relative increase in the intensities of the other lines, such as the OVII-r resonance line, may happen at altitudes on the order of or below 150~km, below the Martian exobase. Those regions should be able to generate oxygen triplet G values much lower than the low density value G$\sim$6. (Similar calculations for the quenching of the O$^{+6}$ metastable state by ionospheric electrons show that this phenomenon is negligible).

Interpretation of the XMM-Newton results in terms of quenching effects requires that the main part of the disk emission as defined in \cite{Dennerl2006} is generated at an altitude of, or below, 150~km. Although our simulations do not include the quenching effect, we can estimate where the bulk of the disk emission comes from. In the $\phi$=40$^{\circ}$ projection presented here, this is not a trivial calculation, since some of the brightest spots on the planet's disk are actually a sum of the line-of-sight (LOS) through several layers of the Martian atmosphere. Therefore, we estimated the contribution of the layer 0-150~km at the limb of the planet in a projection perpendicular to the SW velocity vector, knowing that along this direction, the LOS is tangent to the most emissive parts of the Martian atmosphere in the subsolar point. With this approach we found that the 0-150~km layer contributes no more than 15\% of the total emission between 0-2180~km (the $10''$ radius limit of the disk region). However, the critical density altitude of 150~km is close to our simulation grid resolution, and therefore specific simulations with finer resolution, including the quenching effect, are needed to study this hypothesis in detail.

\section{Discussion and conclusions}\label{discuss}
We have modeled the SWCX X-ray emission produced in the Martian exosphere in SW conditions very close to the ones occurring during the 2003 XMM-Newton observation of Mars \citep{Dennerl2006}. We calculated projected maps of the total SWCX power flux between 0.3 and 0.9~keV, which clearly show the bow shock formed at the SW plasma - planetary neutral interaction region. Just as in modeling of the SWCX emission from the Earth's magnetosheath \citep{Robertson2006}, our model maps demonstrate the potential of SWCX X-ray emission in the studies of electromagnetic structures and SW-neutral interactions around solar system bodies, in support of proposed missions of X-ray imagers \citep{Branduardi2011,Kuntz2008}. 

We find a total luminosity of 0.23~MW in the energy range between 0.3 and 0.9~keV, almost two orders of magnitude lower than the XMM-Newton data (11.8~MW), but closer to the Chandra observations of the planet \citep[0.5$\pm$0.2~MW in the range 0.5-1.2~keV,][]{Dennerl2002a}. The bulk of the emission in our results is found at distances up to 4~R$_M$, compared to 3 and 8~R$_M$ for the Chandra and XMM-Newton data, respectively. 

Previous simulations from \cite{Holmstrom2001} with similar SW conditions (V$_{sw}$ = 400~km~s$^{-1}$, n$_{sw}$ = 2.5 cm$^{-3}$) yielded 2.4 MW (1.52 10$^{25}$ eV/s) for solar minimum and 1.5~MW (0.93 10$^{25}$ eV/s) for solar maximum conditions of the neutral coronas, within a 10 R$_M$ radius. On the other hand, \cite{Gunell2004} find 1.8 MW for a 6~R$_M$ simulation box (with V$_{sw}$ = 330~km~s$^{-1}$, n$_{sw}$ = 4.4 cm$^{-3}$ inputs). Our simulation box only extends up to a $43''$ radius ($\sim$7~R$_M$, see bottom of Figure~\ref{FigAllMap}), but a crude upper limit estimate of the additional emission in the annulus between $43''$ and $50''$ ($\sim$8~R$_M$) only yields an additional 5\% compared to the total luminosity in the box.

Besides the small differences in the proton fluxes used and the size of the simulation boxes, two main effects can account for the divergence of our results from previous simulations. First, all previous simulations have assumed average slow SW abundances for the minor ions. These abundances are from three to five times higher than the real-time data we used in our simulations (see Table~\ref{table:1}) and should have a significant effect (approximately of the same factor) in the comparison with previous simulations. If we apply an average abundance three times higher for all the ions included in our simulations, then the total luminosity reaches $\sim$0.7~MW in the simulation box.

A second effect that may, in part, explain the discrepancies with previous simulations is the lack of the H$_2$ and He coronas in our simulations. According to \cite{Fox2009} and \cite{Krasnopolsky2002,Krasnopolsky2010}, the H$_2$ and He density in the Martian atmosphere can be comparable to the H density, depending on solar conditions. For solar maximum, the H$_2$ density is almost an order of magnitude higher than H at 400~km altitude (but still an order of magnitude lower than O), while He can be anywhere between two and ten times the H density at the same altitude \citep[see profiles extracted from][in Figure~\ref{FigNcor}]{Fox2009,Krasnopolsky2010}. Since the H$_2$ and He corona is not included in our simulation, this may have an important effect on the SWCX luminosity, especially in the disk region. From \cite{Greenwood2001} we deduce that the CX cross-section of highly charged C, N, and O ions with H$_2$ is near to the cross-section for H. Therefore, based on the crude assumption that the SWCX emission is proportional to the neutral density, and had we included the H$_2$ and He coronas in our simulations, the SWCX luminosity could be, at best, two or three times as strong in the disk (assuming that the H$_2$ and He sum to 20 times the H density in the disk). However, the H$_2$ and He coronas are dominant only at low altitudes, up to about 1000-2000~km. After that, the H$_2$ and He densities decrease more rapidly than the H density, and have much less effect on the halo emission. In future work, we intend to include the H$_2$ and He coronas and perform a parametric study of the solar activity influence on the SWCX emission from Mars.

If we sum up the estimated additional emission from the missing H$_2$ and He coronas in the disk and the 5\% of the outer halo region between 7 and 8~R$_M$, the total luminosity in our simulations could be on the order of 0.4~MW under the same SW conditions. If we also adjust to the average slow SW ion abundances, then our simulated luminosity can reach between 1.2 and 2.0~MW, which brings us to almost perfect agreement with previous simulations \citep{Krasnopolsky2000,Holmstrom2001,Gunell2004}, but still a far way from the observed values. 

Another interesting hypothesis for the extended SWCX halo around Mars observed by XMM could be the presence of a hot population of neutral atoms at large distances from the planet. Indeed, according to \cite{Delva2011}, upstream proton cyclotron waves have been observed by up to 11~R$_M$. These waves are generated by pick-up of planetary neutrals by the solar wind, and therefore their occurrence at large distances may come from the presence of an extended hydrogen exosphere around the planet. The large extent of this hot neutral component is further supported by the fact that the excitation of these waves to a measurable amplitude requires an accumulation of ions (generated by ionization of planetary neutrals) in a plasma cell, while it is convected by the SW. Therefore these ions need to be generated by a large quantity of neutrals located upstream in the SW with respect to the position where the cyclotron waves are actually observed. The influence of this hot neutral corona to the ion feedback and electromagnetic interface around the planet has not yet been included in any model of the SW - neutral interaction of the Martian environment to our knowledge, and should be explored in the future.

An additional explanation for the discrepancies between the simulation results and the XMM-Newton data is the large uncertainty of extrapolating L1-based in-situ SW measurements to the Mars orbit. If we consider the longitudinal difference $\Delta \theta \sim 30^{\circ}$ in the positions of Earth and Mars, then the time delay due to the radial propagation of SW particles is somewhat compensated for by the $\sim$27-day solar rotation ($\Delta t = \frac{\Delta r}{V_{sw}} - \frac{\Delta \theta}{13.3^{\circ}/day}$), since Mars is behind the Earth during this period. In that case, the ACE SW window is situated approximately between 2003/11/20 14:00~UT and 2003/11/21 06:00~UT. In this interval the ACE data show a strong density and abundance enhancement related to a halo coronal mass ejection (CME) emitted on 2003/11/18. The SW ion flux N$_i$ measured in this interval was on average 18 times higher than the real-time values we used in our simulations, which could yield a total luminosity of 4~MW ($\sim$6.3~MW, if we include the H$_2$ and He coronas as well), in better agreement with, but still quite lower than, the observed $\sim$12~MW.

\begin{figure}
   \centering
   \includegraphics[width=\linewidth]{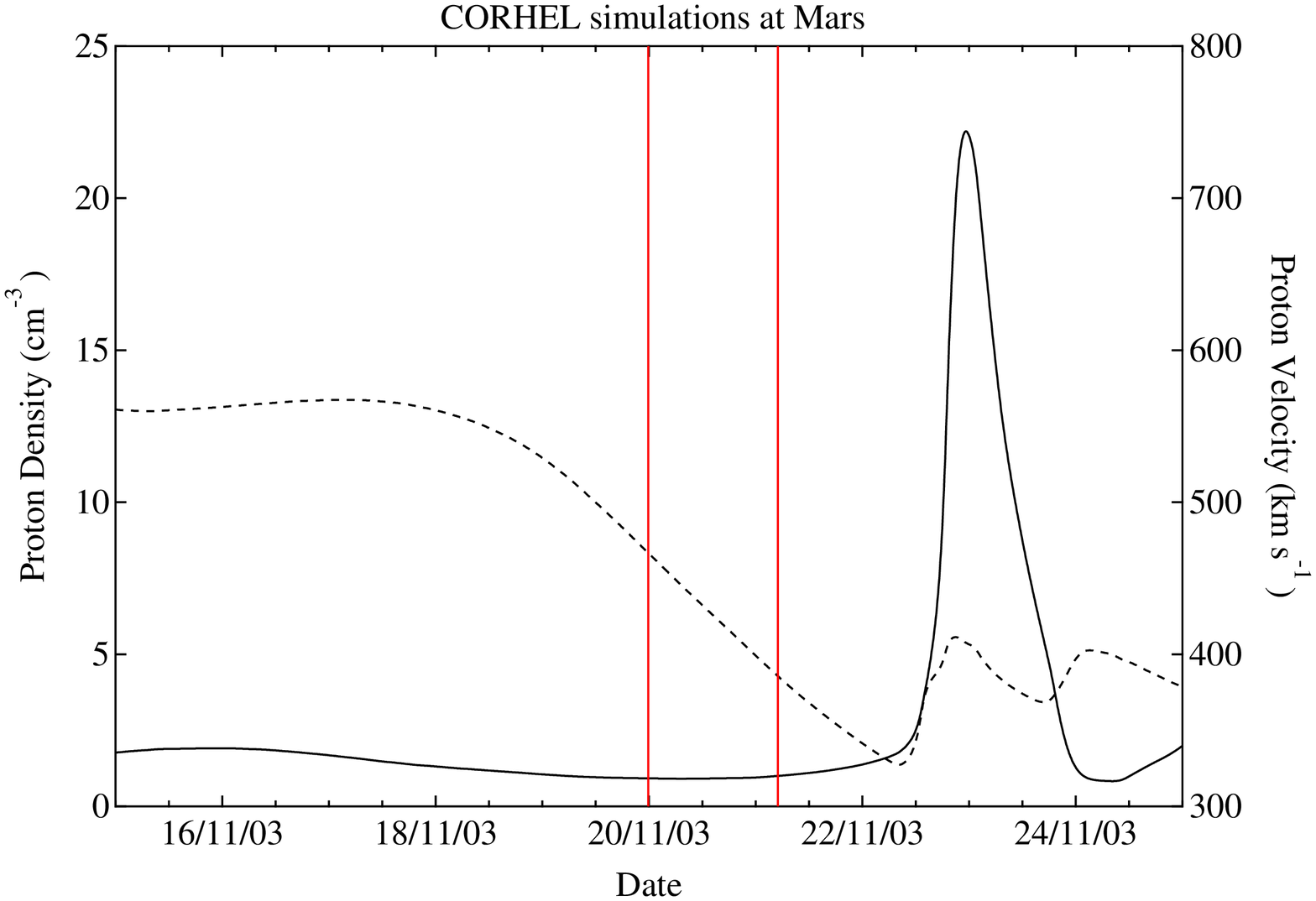}
   \includegraphics[width=\linewidth]{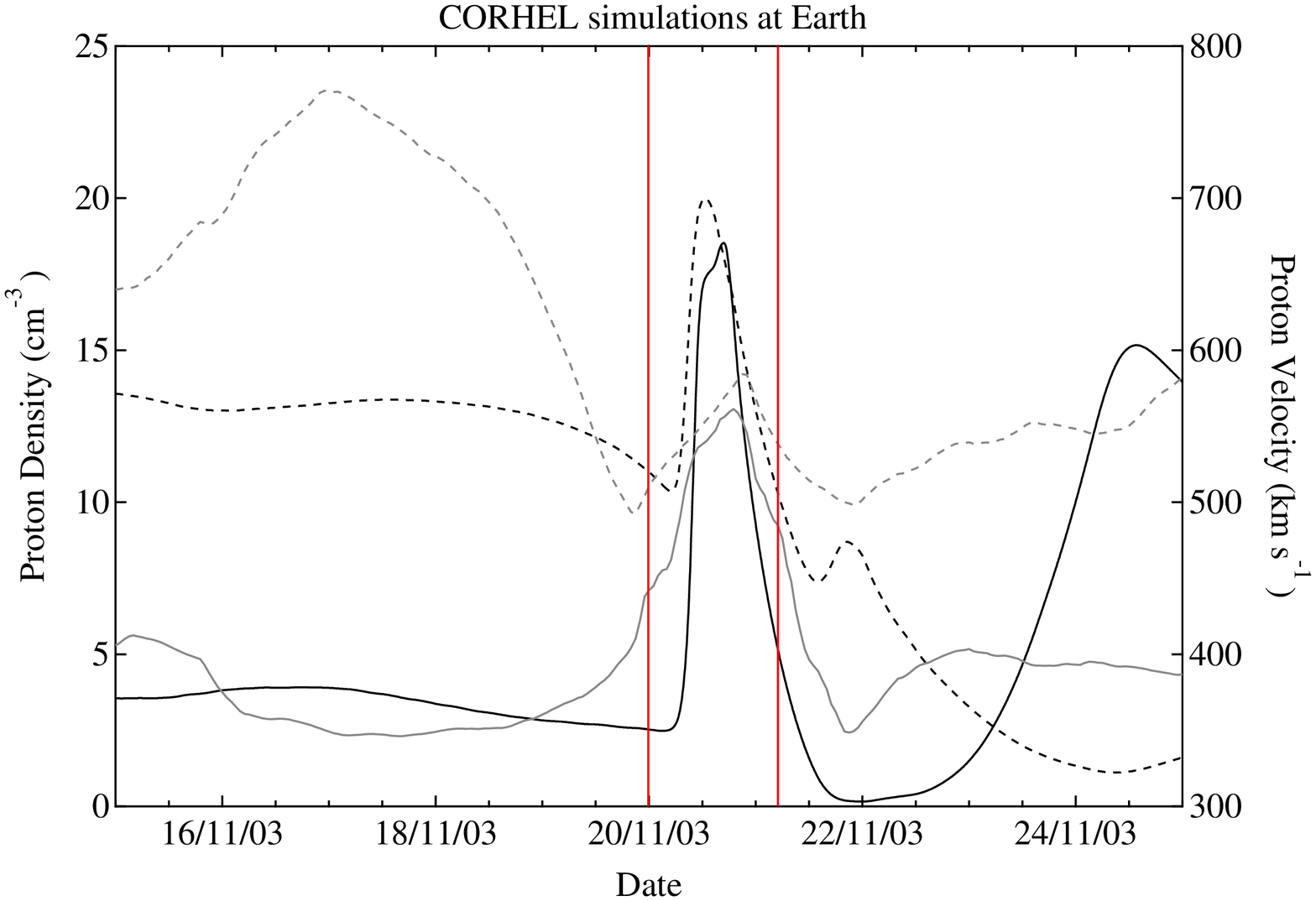}
      \caption{\textit{Top - } Simulated proton density (solid) and velocity (dashed) calculated with the CORHEL model for the Mars orbit for the2003/11/18 CME propagating through interplanetary space. According to the simulations the perturbed CME plasma arrived at Mars about a day and a half after the XMM-Newton exposure (represented by the vertical lines). \textit{Bottom - } Same as top only calculated for the Earth orbit. ACE 1-day gliding averages of density and velocity measurements are also plotted in gray for comparison.}
         \label{FigCORHEL}
\end{figure}

However, a different proof argues against this hypothesis. In Figure~\ref{FigCORHEL} we present results from a CORHEL simulation relevant to the 2003/11/18 CME propagation in interplanetary space. The simulation results were extracted from the Community Coordinated Modeling Center (CCMC \footnote{http://ccmc.gsfc.nasa.gov}) Runs-on-Request system. The top panel of Figure~\ref{FigCORHEL} presents the simulation results for the CME impact on Mars, and we clearly see that the CME arrived a day and a half later than the XMM-Newton exposure window.

To test the timing accuracy of the CORHEL simulations we also checked the simulation results on Earth's orbit and compared them to the ACE data (Figure~\ref{FigCORHEL}-bottom). Although the absolute velocity values in the simulations have up to 200~km~s$^{-1}$ difference compared to the actual data measured by ACE at the L1 point, the timing seems accurate, and so it seems to indicate that the extrapolation to Mars is also accurate and that the CME missed the XMM-Newton observation window at Mars indeed. Even if we apply the data-model velocity difference to estimate the error-margin of the CORHEL simulation, the CME reaches Mars just at the end of the XMM exposure, still too late to effectively influence the SWCX emissivity of the planet. Based on these tests, and owing to the lack of in-situ plasma measurements at Mars simultaneous to the XMM-Newton observations, it is impossible to assert the actual impact of the CME on the planet's SWCX emission. 

For the various reasons discussed above, we feel that the controversy over the very strong SWCX emission detected at Mars with XMM-Newton still remains, especially since it has not been verified by subsequent X-ray observations of the planet. However, all subsequent observations of the planet until now were scheduled during a peculiar solar minimum, when solar activity and SW flux were particularly low. The issue should hopefully be elucidated if a similar event is reproduced. The conditions for such an event are becoming favorable at present and as we approach the solar maximum in the next couple of years, so new X-ray observations of Mars are needed in order to constrain the SWCX emission of the planet.

The discrepancy between the model and the XMM data cannot be lifted through the present analysis, but the sensitivity of the model results to the neutral corona parameters shows that X-ray observations combined with simulations allow the Martian and planetary exospheres, in general, to be studied remotely. This can potentially have interesting applications for studying extrasolar planet exospheres and their mass-loss processes.

Several recent studies report the presence of extended hydrogen atmospheres (as well as C and O, in some cases) around extrasolar planets, revealed by means of absorption of the stellar H~I Lyman-$\alpha$ (C~II and O~I, respectively, for the other components) line when the planet transits in front of the star \citep[e.g.,][]{Vidal-Madjar2003,Ehrenreich2008,LecavelierdesEtangs2010}. An interesting debate arose following the H detection, with the first interpretation supporting the idea that H atoms are escaping the planet's exosphere, owing to hydrodynamic blow-off, and accelerated by stellar radiation pressure \citep{Vidal-Madjar2003,Vidal-Madjar2004, LecavelierdesEtangs2008}. A different approach introduced by \cite{Holmstrom2008}, and further explored by \cite{Guo2011} and \cite{Lammer2011}, proposed that radiation pressure alone cannot suffice and the Lyman-$\alpha$ absorption could be explained by energetic neutral atoms (ENAs) produced in charge exchange collisions between the exospheric neutrals and the stellar wind protons. In this case, ENAs would allow the study of the plasma properties in the stellar wind, and while they do not exclude a large atmospheric escape, they cannot provide any strong constraint on it \citep[see reply by Holmstr\"{o}m et al. to ][]{LecavelierdesEtangs2008}.

If the star is endowed with a solar-like wind, then CX collisions are bound to occur between the  ionized wind particles and exospheric neutrals. In that case, CX-induced X-rays will also be produced if the wind's high ion content is at least solar-like. Therefore, in principle, it would be possible to study the interaction of the stellar wind with the planet's exosphere by CX X-ray imaging and spectroscopy. This would provide a complementary method, less ambiguous than Lyman-$\alpha$ absorption, to study both the plasma and neutral properties of the planet's environment. In theory, the CX X-ray spectroscopy would help to put constraints to the stellar wind ion mass flux and neutral composition and density of the exosphere, since the CX line ratios depend on the collision energy \citep{Kharchenko2003} and the CX cross sections depend on the neutral target \citep{Greenwood2001}. In addition, CX X-ray imaging could constrain the magnetic boundaries around the planet, giving information on whether the planet is strongly magnetized or not.

Although the study of star-planet interaction with CX-induced X-ray imaging and spectroscopy seems promising, we feel that technically it is still unfeasible, as it requires instruments with large collecting areas, and enough spatial and spectral resolution to separate the planet's X-ray signal from the star's. \cite{Wargelin2001} advanced this CX X-ray based approach to study the mass-loss rate of late-type dwarfs through the CX interaction of their winds with neutral material of the IS medium (in the same context as heliospheric SWCX emission), but did not detect any significant CX halo in the cases of Proxima Centauri and the M dwarf Ross 154 studied \citep[][]{Wargelin2002,Wargelin2008}. Recent studies of star-planet interactions in X-rays have mainly focused on the star's X-ray environment to determine the irradiation-driven mass loss of super-Earths or the effects of giant planets on the chromospheric and coronal activity \citep[][]{Poppenhaeger2011,Poppenhaeger2012}, concluding that ``observational evidence of star-planet interaction in X-rays remains challenging''. Hopefully, the next-generation X-ray observatories proposed (e.g., ATHENA) will be able to shed more light in star-planet interactions in extrasolar systems.

\begin{acknowledgements}
We wish to thank our referee Thomas Cravens for his pertinent remarks and our editor Tristan Guillot for stimulating the discussion on exoplanets. 

We are grateful to the ACE SWICS and SWEPAM instrument teams and the ACE Science Center for providing the ACE data. 

The CORHEL simulation results have been provided by the CCMC at Goddard Space Flight Center through their public Runs on Request system (http://ccmc.gsfc.nasa.gov). The CCMC is a multi-agency partnership between NASA, AFMC, AFOSR, AFRL, AFWA, NOAA, NSF and ONR. The CORHEL Model was developed by J. Linker, Z. Mikic, R. Lionello, P. Riley, N. Arge, and D. Odstrcil at PSI, AFRL, U.Colorado.

RM and GC acknowledge financial support for their activity by the program ``Soleil Heliosph\`ere Magn\'etosph\`ere" of the French space agency CNES. RM also wishes to acknowledge Agence Nationale de la Recherche for a supporting grant ANR-09-BLAN-223.
\end{acknowledgements}





      
\bibliographystyle{aa} 
\bibliography{BiblioDK.bib} 
\end{document}